\documentstyle[12pt]{l-aa} 
\input epsf 

\def\gsim{\lower.4ex\hbox{$\;\buildrel >\over{\scriptstyle\sim}\;$}}
\def\lsim{\lower.4ex\hbox{$\;\buildrel <\over{\scriptstyle\sim}\;$}}

  \def\bib{\bibitem{}}
\newcommand{\xia}{\overline{\xi}}
\newcommand{\rhoa}{\overline{\rho}}

\newcommand{\rhot}{\tilde{\rho}}
\newcommand{\betat}{\tilde{\beta}}
\newcommand{\gam}{\gamma}
\newcommand{\real}{\bbbr}
\newcommand{\natur}{\bbbn}
\newcommand{\naturs}{\bbbn^*}
\newcommand{\Pa}{\overline{P}}
\newcommand{\om}{\omega}

\newcommand{\inta}{\int_{-i\infty}^{+i\infty}}
\newcommand{\beq}{\begin{equation}}
\newcommand{\eeq}{\end{equation}}
\newcommand{\lag}{\langle}
\newcommand{\rag}{\rangle}
\begin{document}
%
% US additions
% 
\topmargin=2.5 cm

\thesaurus{Sect.02 (12.12.1; 11.03.1)} 
\title{Non-linear gravitational clustering: smooth halos, substructures and scaling exponents.}
\author{Patrick Valageas}
\institute{Service de Physique Th\'eorique, CEA Saclay, 91191 
Gif-sur-Yvette, France}
\date{Received ;  }
\maketitle 
\markboth{P. Valageas: Non-linear gravitational clustering: smooth halos, substructures and scaling exponents.}{P. Valageas: Non-linear gravitational clustering: smooth halos, substructures and scaling exponents.}

\begin{abstract}

Within the framework of hierarchical clustering scenarios, we investigate the consequences for the properties of virialized halos of the constraints provided by numerical simulations on the first few correlation functions. Thus, we show that the density field {\it cannot be described by a collection of smooth halos} with a universal density profile. This implies that substructures within larger objects play an important role (but a mean spherically averaged density profile may exist). In particular, a possible interpretation is that collapsed objects can be divided into an infinite hierarchy of smaller objects with increasingly large densities (these substructures might also be continuously destroyed and created by the long-range action of  gravity). Finally, we present multifractal models (restricted to non-linear scales) which can describe in a natural way such non-linear density fields with increasingly large fluctuations at smaller scales. We relate their properties to the correlation functions and present a few constraints they are expected to satisfy, using theoretical considerations as well as constraints from numerical simulations. Thus, the simplest realistic model is the {\it bifractal} model described in Balian \& Schaeffer (1989a). Moreover, we show that it should provide (at least) a very good approximation of the multifractal properties of the actual non-linear density field, hence of the probability distribution of the density contrast. The implications of this model (e.g. for galaxies) are detailed in other studies.

\end{abstract}


\section{Introduction}

In the standard cosmological scenario the non-linear gravitational structures we observe in the present universe have formed by the amplification of small primordial density fluctuations. Moreover, according to hierarchical models the power increases at small scales so that low masses collapse first to build small virialized objects which will later merge to form broader halos as larger scales turn non-linear. The objects built by this continuous process produce the galaxies and clusters we observe today. Thus, the description of the non-linear evolution of the density field is an important task in cosmology. Most works have considered three tools to measure its properties. First, one can study the mass function of halos defined for instance by a given density threshold. This is the goal of the popular Press-Schechter approximation (Press \& Schechter 1974). Note however that this analytical approach is restricted to only one value of the density contrast threshold ($\Delta \sim 177$). Second, one can consider the density profile of individual objects. This is mainly done through numerical simulations (e.g. Navarro et al.1996; Tormen et al.1997). Third, one can measure the many-body correlation functions. These are closely related to the counts-in-cells: the universe (or the simulation box) is divided into cells of a given size and one considers the probability distribution of the density contrast realized in such a cell. This can be studied through numerical simulations (e.g. Colombi et al.1997; Valageas et al.1999a) and analytical works based on specific scaling models (Balian \& Schaeffer 1989a).

While the first two properties are directly linked to the characteristics of the astrophysical objects we observe the third has the advantage to be better defined (it bypasses the problem of the recognition of individual halos) and to contain much more information. In particular, it should be possible to obtain the properties of individual halos from the correlation functions. For instance, the mass functions of objects defined by {\it various} density thresholds (not necessarily constant) obtained from the scaling model developped by Balian \& Schaeffer (1989a) for the correlation functions were considered in Valageas \& Schaeffer (1997) and Valageas et al.(1999a). 

In this article, we investigate the general link between the behaviour of the correlation functions and the shape of collapsed halos. First, in Sect.\ref{Scaling exponents} we introduce our notations. We relate the correlation functions to the moments of the density probability distribution and we note a few exact properties. In Sect.\ref{Constraints} we recall the constraints on the first few correlation functions provided by numerical simulations. Then, in Sect.\ref{Single power-law profile} and Sect.\ref{Mass-dependent profile} we show that the density field {\it cannot} be described as a collection of {\it smooth} halos with a universal profile. This points out the role of substructures, as discussed in Sect.\ref{Substructures}. Finally, in Sect.\ref{Multifractals} we present multifractal models which provide a natural tool to describe a density field with increasingly large density fluctuations at smaller scales. We discuss their link with correlation functions and we obtain some constraints from theoretical considerations as well as from numerical simulations. In particular, we show that a bifractal model should provide (at least) a very good approximation for the non-linear density field.

\section{Scaling exponents}
\label{Scaling exponents}

A convenient description of the density field is provided by the
probability distribution $P_l(\rhot_l) \; d\rhot_l$ of the local mean
overdensity over spheres of radius $l$ and volume $V_l$. Here we
defined: 
\beq 
\rhot({\bf r}) = \frac{\rho({\bf r})}{\rhoa} =
(1+\delta)({\bf r}) 
\eeq 
and 
\beq 
\rhot_l({\bf r}) = \int_{V_l}
\frac{d^3 r'}{V_l} \rhot({\bf r'})  \hspace{0.5cm} > 0
\eeq 
Since usual probability distributions can be defined by their moments
one is led to consider the parameters $\mu_p(l)$ given by: 
\beq
p \in \real \; : \hspace{0.5cm} \mu_p(l) = \lag \rhot_l^{\;p} \rag = \int_0^{\infty} d\rhot \; \rhot^{\;p} \; P_l(\rhot)
\label{defmu} 
\eeq 
where $\lag \rag$ denotes an average over the position ${\bf r}$ of the
sphere. Of course, if all moments $\mu_p$ ($p \in \real$) are finite their values for $p \in \natur$ (or $-p \in \natur$) are sufficient to define the density field. Hence all parameters $\mu_p$ ($p \in \real$) can be obtained from the sole $\mu_p$ realized for $p \in \natur$ (or $-p \in \natur$). From the definition (\ref{defmu}) one can easily show that:
\beq
p \leq p' \; , \; q \geq 0 \; : \; \frac{\mu_{p+q}}{\mu_p} \leq \frac{\mu_{p'+q}}{\mu_{p'}}
\label{mupq}
\eeq
In particular, we obtain: 
\beq 
p \in \real \; : \hspace{0.5cm} \frac{ \mu_{p+1} }{ \mu_p } \geq \frac{ \mu_p }{
\mu_{p-1} } 
\label{mup1}
\eeq 
Since $\mu_1=1$ we have for positive integers: 
\beq 
p \in \naturs \; : \hspace{0.5cm} \frac{ \mu_{p+1} }{ \mu_p } \geq \mu_2 \hspace{0.5cm} \mbox{and} \hspace{0.5cm} \mu_p \geq \mu_2^{p-1} 
\label{mupmu2}
\eeq 
We also obtain, using $\mu_0=1$ (and the fact that $(1+\delta)$ and
$(1+\delta)^{-1}$ are not proportional), 
\beq 
p \in \naturs \; : \hspace{0.5cm}  \frac{ \mu_{-(p+1)} }{ \mu_{-p} } \geq 
\frac{ \mu_{-p}}{ \mu_{-(p-1)} } \geq \mu_{-1} > 1
\label{mump1} 
\eeq 
hence: 
\beq 
p \in \naturs \; : \hspace{0.5cm} \mu_{-p} \geq \mu_{-1}^p 
\eeq 
From the definition (\ref{defmu}) we can also write: 
\beq
p \in \naturs \; : \hspace{0.5cm} \mu_p(l) = \left< \int_{V_l} \frac{d^3r_1... d^3r_p}{V_l^p} \rhot({\bf r}_1) ... \rhot({\bf r}_p) \right>
\label{muxi}
\eeq 
which emphasizes that these moments are related to the $p-$point
correlation functions $\xi_p$. We define: 
\beq 
\xia_p(l) = \int_{V_l} \frac{d^3r_1... d^3r_p}{V_l^p} \; \xi_p ({\bf
r}_1,...,{\bf r}_p) \hspace{0.3cm} , \hspace{0.3cm} \xia_0=\xia_1=1
\eeq 
Then, from (\ref{muxi}) and the definition of correlation functions (Peebles 1980) we have:
\beq
p \in \naturs \; : \;\; \mu_p(l) = \sum_{i_1,..,i_p} \; a_{i_1,..,i_p} \; \xia_{i_1}(l) ... \xia_{i_p}(l)
\label{ai}
\eeq
where the sum runs over the integers $0 \leq i_1 \leq ... \leq i_p$ with $i_1+...+i_p=p$, the coefficients $a_{i_1,..,i_p}$ are positive and $a_{1,..,1} = a_{0,..,0,p} = 1$. In particular,
\beq
\mu_2 = 1 + \xia
\eeq
where we note $\xia_2$ as $\xia$. As a consequence, in the non-linear regime we obtain from (\ref{mupq}) and (\ref{ai}):
\beq
p \in \natur \; , \; \xia \rightarrow \infty \; : \hspace{0.5cm} \frac{\mu_p}{\xia_p} \rightarrow 1
\label{mupxip}
\eeq
Of course this also implies that the density field is strongly non-gaussian in the non-linear regime. We are interested in the case of a critical universe with an initial power-spectrum which is a power-law: $P(k) \propto k^n$. Then, it is natural to expect a power-law behaviour for the various quantities $\xia_p(l)$ and $\mu_p(l)$ in the non-linear regime $\xia \gg 1$ (i.e. at small scales). Note indeed that numerical simulations show a power-law behaviour for $\xia$ (Jain et al.1995; Colombi et al.1996; Valageas et al.1999a). Thus, we define the limits:
\beq
p \in \real \; : \hspace{0.5cm} \gam_p = - \lim_{l \rightarrow 0} \frac{\ln \mu_p(l)}{\ln l}
\label{gampl}
\eeq
From (\ref{mupxip}) this means that for integers $p$ we have:
\beq
p \in \natur \; : \hspace{0.5cm} \xia_p(l) \sim \l^{-\gam_p} \hspace{0.5cm} 
\mbox{for} \hspace{0.5cm} l \rightarrow 0
\eeq
where we ignore logarithmic corrections. From (\ref{mup1}) and (\ref{mump1}) we obtain:
\beq
p \in \real \; : \hspace{0.5cm} \left\{ \begin{array}{l} \gam_{p+1} - \gam_p \geq \gam_p - \gam_{p-1} \\  \\ \gam_{-(p+1)} - \gam_{-p} \geq \gam_{-p} - \gam_{-(p-1)} \end{array} \right.
\label{gamp1}
\eeq
In particular, we have:
\beq
\gam_0=\gam_1=0 \; , \hspace{0.5cm} p \in \naturs \; : \; \left\{ \begin{array}{l} \gam_p \geq (p-1) \; \gam_2 \\  \\  \gam_{-p} \geq p \; \gam_{-1} \end{array} \right.
\label{zetap2}
\eeq
Moreover, since $\rhot=(1+\delta)$ is a positive quantity $l^3 \rhot_l({\bf r})$ is a non-decreasing function of $l$ for any point ${\bf r}$. This implies that for any $p \geq 0$ the average $l^{3 p} \lag \rhot_l^{\;p} \rag$ is a non-decreasing function of $l$, so we get: 
\beq
p \geq 0 \; : \hspace{0.5cm} 3 p - \gam_p \geq 0 \hspace{0.5cm} \mbox{hence} \hspace{0.5cm} \lim_{p \rightarrow \infty} \frac{\gam_p}{p} \leq 3
\label{gamp3}
\eeq
Note that all the results obtained above are only due to the fact that $\rhot \geq 0$. To obtain the probability distribution $P_l(\rhot) \; d\rhot$ of the overdensity at scale $l$ it is convenient to introduce the quantities:
\beq
p \in \naturs \; : \hspace{0.5cm} S_p(l) = \frac{\xia_p(l)}{\xia(l)^{p-1}} \hspace{0.5cm} , \hspace{0.5cm} 
S_1=1
\label{Sp}
\eeq
and to define the function:
\beq           
\varphi_l(y)   =   \sum_{p=1}^{\infty} \frac{(-1)^{p-1}}{p!} \; S_p(l) \; y^p 
\label{phiy}
\eeq
Then, one can show (White 1979; Balian \& Schaeffer 1989a) that:
\beq
P_l(\rhot) = \inta \frac{dy}{2\pi i \xia} \; e^{(\rhot y - \varphi_l(y))/\xia}
\label{Pl}
\eeq
The interest of the function $\varphi_l(y)$ is that it shows a very small (if any) dependence on scale $l$ in the non-linear regime, so that the evolution with $l$ of the probability distribution $P_l(\rhot)$ is given by the sole two-point correlation function $\xia$. This was shown in the non-linear regime from numerical simulations by Colombi et al.(1992, 1994, 1995, 1997), Munshi et al.(1999) or Valageas et al.(1999a) for instance. Note that this property also holds in the quasi-linear regime (Bernardeau 1994), with a different function $\varphi_l(y)$. In the regime $\xia \gg 1$ which we consider in this article, we have from (\ref{Sp}) and (\ref{zetap2}) for $p \geq 3$:
\beq
l \rightarrow 0 \; : \hspace{0.3cm} S_p(l) \sim l^{-\zeta_p} \hspace{0.3cm} , \hspace{0.3cm} \zeta_p = \gam_p - (p-1) \; \gam_2 \geq 0
\label{zetap}
\eeq
Thus, {\it the coefficients $S_p$ are constant or increase with $\xia$ } (neglecting logarithmic corrections). We recall in appendix the behaviour of $P_l(\rhot)$ and the moments $\mu_p$ implied by simple forms of $\varphi_l(y)$. Note that one may also define from the moments $\mu_p$ the generating function:
\beq
\psi_l(y) =  \sum_{p=0}^{\infty} \frac{(-1)^{p}}{p!} \; \mu_p(l) \; y^p 
\eeq
which leads to:
\beq
P_l(\rhot) = \inta \frac{dy}{2\pi i} \; e^{\rhot \; y} \; \psi_l(y)
\eeq
and:
\beq
\psi_l(y) = e^{-\varphi_l(\xia \; y)/\xia}
\label{psiphi}
\eeq
Developping (\ref{psiphi}) in $y$ one can obtain the coefficients $a_{i_1,..,i_p}$ defined in (\ref{ai}).

\section{Constraints}
\label{Constraints}

The behaviour of the two-point correlation function $\xia$ and of the first coefficients $S_p$ with $p \leq 6$ has been studied by many numerical works in the non-linear regime. Thus, the slope $\gam_2$ (usually noted $\gam$) of $\xia$ is consistent with the prediction of the stable-clustering ansatz (Davis \& Peebles 1977; Peebles 1980):
\[
{\bf C1} \; : \hspace{0.5cm} \gam = \frac{3(3+n)}{5+n}  
\]
where $n$ is the slope of the initial power-spectrum (Jain et al.1995; Colombi et al.1996; Valageas et al.1999a). Moreover, the stable-clustering assumption predicts that in the highly non-linear regime the $p-$point correlation functions satisfy the scaling law:
\beq
\xi_p(\lambda {\bf r}_1,...,\lambda {\bf r}_p ;a) = a^{3(p-1)} \; 
\lambda^{-\gam(p-1)} \; \hat{\xi}_p({\bf r}_1,...,{\bf r}_p)  \label{scal1}
\eeq
where $a(t)$ is the scale-factor, which means that $\zeta_p=0$ and the coefficients $S_p$ do not depend on the scale $l$. Colombi et al.(1996) found a very small drift $S_p \propto \xia^{\; 0.045 (p-2)}$ for $p \leq 5$ while Munshi et al.(1999) obtained no scale dependence for $p \leq 6$. Thus, numerical simulations provide the conservative constraint:
\[
{\bf C2} \; : \hspace{0.5cm} 0 \leq \zeta_p \leq 0.045 \; (p-2) \; \gam \hspace{0.5cm} \mbox{for } 3 \leq p \leq 5
\]
Hence, the first scaling exponents $\gam_p$ are very close to the stable clustering prediction $\gam_p^{sc}$:
\beq
3 \leq p \leq 5 \; : \hspace{0.5cm}  0 \leq \frac{\gam_p - \gam_p^{sc}} {\gam_p^{sc}} \leq 0.045 \; \frac{p-2}{p-1}
\eeq
\beq
\mbox{with} \hspace{0.5cm} p \geq 1 \; : \hspace{0.5cm} \gam_p^{sc} = (p-1) \; \gam
\label{gampsc}
\eeq
where $\gam$ is given by ${\bf C1}$. The results ${\bf C1}$ and ${\bf C2}$ provided by numerical simulations are certainly robust. Indeed, they are derived (among other methods) from the counts in cells which are a very well defined procedure. By contrast, we think that the properties of dense halos seen in numerical simulations may be more uncertain since it is not always easy to identify these objects which may not be spherical. However, many numerical studies have considered the density profile of virialized halos. In particular, Navarro et al.(1996, 1997) found that the spherically averaged density profile follows a universal behaviour, independent of time and of the initial power-spectrum (although the numerical parameters which enter this average profile depend on the index $n$ of the power-spectrum its shape does not depend on $n$). Thus, many studies (e.g. about galaxy formation processes) assume that:\\

${\bf H1}$ : {\it the density field can be described as a collection of spherical halos with a smooth density profile given by a universal law}.\\

We note that this simple (and convenient) picture neglects substructures within large halos. Thus, ${\bf H1}$ is more restrictive than the assumption of a spherically averaged universal profile which removes substructures by definition. However, the validity (or non-validity) of ${\bf H1}$ will provide some information on the importance of substructures. We note that numerical simulations often suffer from the overmerging problem: due to poor resolution small objects often get disrupted within large halos. Nevertheless, some simulations with a high resolution show significant substructures (Moore et al.1998; Ghigna et al.1998). In the next sections, we shall check whether the description ${\bf H1}$ is consistent with the constraints ${\bf C1}$ and ${\bf C2}$.

\section{Single power-law profile}
\label{Single power-law profile}

We first consider the simplest case where the density field can be described as a collection of spherical halos with a smooth power-law density profile and a fixed overdensity $(1+\Delta_c)$ at their radius $R$ (e.g. $\Delta_c \sim 200$ from the spherical collapse model). We note the multiplicity function of virialized halos of mass $M$, volume $V(M)$, as $\eta(M) \frac{dM}{M}$ while the local overdensity at distance $r$ from the center of a halo of radius $R$ is:
\beq
\rhot_h(r) = (1+\Delta_c) \; \left( \frac{r}{R} \right)^{-\beta}
\eeq
where $0<\beta<3$ is the slope of the density profile. The mass function satisfies:
\beq
\int_0^{\infty} \frac{M}{\rhoa} \eta(M) \frac{dM}{M} = 1
\label{etaM}
\eeq
which states that the fraction of matter enclosed in these halos is unity. It could be smaller if one adds for instance a uniform component. However, this would not change our results since they only rely on the fact that the integral in (\ref{etaM}) is finite and this low density component only contributes by a constant amount to the moments $\mu_p$ ($p \geq 2$) which tend to infinity when $l \rightarrow 0$. We want to obtain the exponents $\gam_p$ for $p \geq 2$ realized in such a density field, to compare them with the constraints ${\bf C1}$ and ${\bf C2}$. Thus, for the sake of clarity we shall not consider the normalization factors which enter our calculations since we only need the exponents of the various power-laws of interest. To obtain the behaviour of $\mu_p$ we must separate large halos ($R>l$: one cell is within one such halo) from small halos ($R<l$: one cell may contain several whole halos). Thus, we write:
\beq
\mu_p(l) \sim A_1(l) + A_2(l)
\eeq
where $A_1$ is the contribution of cells which are enclosed within a massive halo ($R>l$) while $A_2$ is the contribution of cells which contain zero, one or several small objects. Note that for $R \sim l$ our distinction is not adequate but we will not consider in detail such intermediate regimes which do not influence the power-law behaviours we are interested in. Then,
\beq
A_1(l) \sim \int_{R=l}^{\infty} \eta(M) \frac{dM}{M} V(M) \lag\rhot^{\;p}\rag_{l,M}
\eeq
where $\lag\rhot^{\;p}\rag_{l,M}$ is the mean overdensity to the power $p$ seen by a cell $l$ which intersects a halo $M$. Thus,
\beq
\lag\rhot^{\;p}\rag_{l,M} \sim \frac{1}{V} \left[ \int_l^R 4 \pi r^2 dr \rhot_h(r)^p + \frac{4 \pi}{3} l^3 \rhot_h(l)^p \right]
\eeq
which leads to:
\beq
\lag\rhot^{\;p}\rag_{l,M} \sim (1+\Delta_c)^p \; \left[ 1 + \left( \frac{l}{R} \right)^{3-p \beta} \right]
\eeq
As explained above, here the factor $1$ stands for a numerical constant. Thus, we have:
\[
A_1 \sim \int_{R=l}^{\infty} \eta(M) \frac{dM}{M} \frac{M}{(1+\Delta_c)\rhoa} (1+\Delta_c)^p \; \left[ 1 + \left( \frac{l}{R} \right)^{3-p \beta} \right]
\]
Considering separately the cases $(3-p \beta)<0$ and $(3-p \beta)>0$, and using (\ref{etaM}), we can push the lower bound of the integral down to $0$ and we obtain:
\beq
l \rightarrow 0 \; : \hspace{0.5cm} A_1(l) \sim l^0 + l^{3-p \beta}
\eeq
Here, we also assumed that $\int_0^{\infty} M^{1+\alpha} \eta(M) dM/M$ converges for any $\alpha>0$. This means that the multiplicity function has an ``exponential'' cutoff at larges masses which is consistent with numerical results and observations of the galaxy luminosity function (of course the argument of the exponential is not necessarily linear in $M$). This ensures that $\mu_p(l)$ is finite for any $p \geq 2$ for a finite $l>0$. On the other hand, the overdensity within cells which contribute to $A_2(l)$ is smaller than $(1+\Delta_c)$ (which is realized when the whole cell is covered by small halos) which means that $A_2(l)$ has a finite limit for $l \rightarrow 0$. Thus, we obtain:
\beq
p \geq 2 \; : \;\; \mu_p(l) \sim l^0 + l^{3-p \beta}
\eeq
If this density field is consistent with ${\bf C1}$ then:
\beq
\beta = \frac{3+\gam}{2} = \frac{3(n+4)}{n+5} 
\label{beta}
\eeq
as was obtained by McClelland \& Silk (1977), and moreover
\beq
p \geq 2 \; : \hspace{0.4cm} \gam_p = p \beta - 3 \hspace{0.4cm} , \hspace{0.4cm} \zeta_p = (p-2) \; \frac{3-\gam}{2}
\label{betagamp}
\eeq
Thus, the exponents $\zeta_p$ violate the constraint ${\bf C2}$ for any power-spectrum of interest $-3<n<1$. This shows that the model considered in this section cannot provide an accurate description of the actual density field. Moreover, we note that the slope $\beta$ needed to reproduce the two-point correlation function $\xia$ is much larger than the density profile observed in numerical simulations ($\beta=1$ in Navarro et al.1997; $\beta=1.4$ in Moore et al.1998).

\section{Mass-dependent profile}
\label{Mass-dependent profile}

While the density profile of the halos considered in the previous section was given by a unique power-law for all objects, many authors have found that small objects are denser than larger ones. Thus, Navarro et al.(1996) obtain:
\beq
\rhot_h(r) = \frac{\delta_c}{(r/R_s)(1+r/R_s)^2}
\label{rhot}
\eeq
with
\beq
\delta_c = \frac{200}{3} \; \frac{c^3}{\ln(1+c) - c/(1+c)}
\label{dc}
\eeq
where $R_s=R/c$ is a characteristic radius ($c \geq 1$) and small objects are more centrally concentrated: they have a higher $c$ and $\delta_c$. The same trend is seen (Tormen et al.1997) using a Hernquist profile (Hernquist 1990):
\beq
\rhot_h(r) = \frac{\delta_c}{(r/R_s)(1+r/R_s)^3}
\label{Hernquist}
\eeq
Thus, these objects have a mean overdensity $(1+\delta) \sim \delta_c$ over the radius $R_s$ and $(1+\delta) \sim 200$ over $R$. Navarro et al.(1996) have found that the characteristic overdensity of these halos is proportional to the mean density of the universe at the epoch they were formed (which we note by the redshift $z_f$): $\delta_c \propto (1+z_f)^3$. This redshift $z_f$ increases for smaller halos which formed when the corresponding scale $M$ became non-linear. Indeed, using the extended Press-Schechter model (Press \& Schechter 1974) developped by Lacey \& Cole (1993, 1994), one obtains good results by setting:
\beq
\mbox{erfc} \left( \frac{1.69 \; z_f(M)}{\sqrt{2 (\sigma^2(M/2) - \sigma^2(M))}} \right) = \frac{1}{2} 
\label{zf}
\eeq
where as usual $\sigma^2(M)$ is the variance of the density fluctuations at scale $M$ in the present universe given by the linear theory (Lacey \& Cole 1994). Note however that Navarro et al.(1996,1997) get better results by using a small mass fraction $f=0.01$ rather than $f=1/2$ in (\ref{zf}). For small masses, this leads to:
\beq
M \ll M_* \; : \hspace{0.5cm} (1+z_f) \sim  \sigma(M) = \left( \frac{M}{M_*} \right)^{-(n+3)/6}
\eeq
where we defined $M_*$ by $\sigma(M_*)=1$. Hence, we obtain:
\beq
M \ll M_* \; : \hspace{0.5cm} \delta_c \sim \left( \frac{M}{M_*} \right)^{-(n+3)/2}
\label{dcM}
\eeq
As we noticed above, this simply states that the characteristic density of halos of mass $M \ll M_*$ is given by the mean density of the universe at the time when this mass scale became non-linear. This behaviour is probably more robust than the detailed prescription (\ref{zf}). We note that this means there has only been a negligible evolution of these small objects: their density has not significantly changed since their formation. Note that Salvador-Sole et al.(1998) also recovered (\ref{dcM}) using a more detailed model than (\ref{zf}) which takes into account violent mergers and slow accretion. Neglecting the logarithmic correction in (\ref{dc}) we also have:
\beq
M(<R_s) \sim \rhoa \; \delta_c \; R_s^3 \sim \rhoa \; c^3 \; \left( \frac{R}{c} \right)^3 \sim \rhoa \; R^3 \sim M(<R)
\eeq
This means that most (or a finite fraction) of the mass is enclosed within $R_s$. This would be exact for any outer density profile ($r>R_s$) steeper than $\rhot_h(r) \propto r^{-3}$, like the Hernquist profile (\ref{Hernquist}). This is also consistent with the results of Navarro et al.(1996) since simulations cannot distinguish between $\rhot_h(r) \propto r^{-3}$ and $\rhot_h(r) \propto r^{-3.1}$ (for instance). Thus, we shall consider in this section a model where large halos have a single power-law density profile: 
\beq
M > M_* \; : \hspace{0.5cm} \rhot_h(r) = (1+\Delta_c) \; \left( \frac{r}{R} \right)^{-\beta}
\eeq
while small objects follow a double power-law profile:
\beq
M < M_* \; : \hspace{0.5cm} \frac{R_s}{R} = \left( \frac{M}{M_*} \right)^{(n+3)/6} \hspace{0.5cm} \mbox{and}
\eeq
\[
\left\{ \begin{array}{ll} r < R_s \; : & \rhot_h(r) = (1+\Delta_c) \; \left( \frac{M}{M_*} \right)^{-(n+3)/2} \; \left( \frac{r}{R_s} \right)^{-\beta}  \\   \\  R_s < r < R \; : & \rhot_h(r) = (1+\Delta_c) \; \left( \frac{M}{M_*} \right)^{-(n+3)/2} \; \left( \frac{r}{R_s} \right)^{-\betat}  \end{array} \right.
\]
with:
\beq
\beta < 3  \hspace{0.5cm} , \hspace{0.5cm} \betat > 3
\eeq
Of course, this double power-law profile is a simplification of the smooth profiles (\ref{rhot}) and (\ref{Hernquist}). However, it captures their main features which are that a finite fraction of the mass is enclosed within a radius $R_s$ with an inner density profile with a slope $\beta=1$. The outer slope $\betat$ will not enter our final results (it only ensures that $R_s$ contains a finite fraction of the mass and that $M \propto R^3$). Since small halos may provide a significant contribution to the moments $\mu_p$ (contrary to the case considered in the previous section) because of their higher density, we must detail the behaviour of the multiplicity function in the low mass limit. Thus we write:
\beq
M \ll M_* \; : \hspace{0.5cm} \eta(M) \propto M^{\theta-1} \hspace{0.5cm} , \hspace{0.5cm} \theta > 0
\label{alp}
\eeq
where we did not consider logarithmic corrections. The lower bound $\theta > 0$ is given by the condition (\ref{etaM}). In most models, one also has $\theta<1$ so that the number of small objects diverges (as in the Press-Schechter mass function, Press \& Schechter 1974). We note that this is also necessary to reproduce the faint-end slope of the galaxy luminosity function, since the luminosity usually scales as a power of the dark matter halo mass in order to obtain the Tully-Fisher relation (although the actual power-spectrum is probably not a power-law it is smooth enough to be approximated by a power-law over the mass range corresponding to galaxies). However, we shall not need this upper bound in the following. The fraction of matter (or luminosity) contributed by small objects in these models and observations also decreases for small $M$ (or $L$) as a power-law with a positive exponent, that is $\theta > 0$. Thus we do not consider here the case $\theta=0$ which can still be made consistent with (\ref{etaM}) with suitable logarithmic prefactors (e.g. $\eta(M) \propto 1/[M \ln^2(M)]$). Hence, this model corresponds to the hypothesis ${\bf H1}$ where the profile of the halos is given by the results of Navarro et al.(1996) or similar studies (Hernquist 1990; Moore et al.1998). In a fashion similar to the calculations shown in the previous section we can obtain the scale dependence of the moments $\mu_p$ for $p \geq 2$. Thus, we write:
\beq
\mu_p(l) \sim A_1(l) + A_2(l) + A_3(l) + A_4(l)
\eeq
where $A_1$ corresponds to cells embedded within a large halo $M>M_*$, $A_2$ to cells enclosed within an object $M<M_*$ such that $l<R_s$, $A_3$ to cells within a halo $M<M_*$ with $R_s<l<R$ and $A_4$ to cells which contain zero, one or several small objects. Indeed, since we are interested in the small scale behaviour we have $l \ll R_*$ (where $R_*$ defined by $\sigma(R_*)=1$ is of the order of the radius of collapsed halos of mass $M_*$). We obtain: 
\[
A_1 \sim l^0 + l^{3-p \beta} \hspace{0.3cm} , \hspace{0.3cm}  A_2 \sim l^0 + 
l^{3-p \beta} + l^{\frac{6}{n+5} [\theta - (p-1) (n+3)/2]} 
\]
\[
A_3 \sim l^{\frac{6}{n+5} [\theta - (p-1) (n+3)/2]} \hspace{0.3cm} , \hspace{0.3cm} A_4 \; \mbox{finite}
\]
Hence, the exponents $\gam_p$ are given by:
\beq
p \geq 2 \; : \hspace{0.3cm} \gam_p = \mbox{Max} \left[ p \beta -3 , (p-1) \frac{3(3+n)}{5+n} - \frac{6 \theta}{5+n} \right]
\label{gampalp}
\eeq
Since $\theta>0$, the constraint ${\bf C1}$ implies (\ref{beta}) and (\ref{betagamp}). Thus, we recover the results of the previous section. This means that this model cannot provide a satisfactory description of the density field either.

\section{Substructures}
\label{Substructures}

The results of Sect.\ref{Single power-law profile} and Sect.\ref{Mass-dependent profile} show that the assumption ${\bf H1}$ is not valid: one cannot describe the dark matter density field by a collection of halos with a smooth density profile. In other words, {\it the substructures embedded within larger halos play an important role}: they have to be taken into account in order to obtain the correct scaling exponents $\gam_p$. This is also directly seen from (\ref{gampalp}). Indeed, we note that both constraints ${\bf C1}$ and ${\bf C2}$ are satisfied if $\theta=0$. As we noticed in the previous section, this cannot be used as a satisfactory mass function of distinct halos (even after addition of ad-hoc logarithmic factors to ensure convergence) because this would contradict the observed galaxy luminosity function and the mass functions measured in numerical simulations. However, the value $\theta=0$ has a very simple and natural interpretation: {\it one counts the same matter (particles) at small scales as is counted at larger scales}. In other words there is some multiple counting and the same mass is seen alternatively as objects of scale $M$ and smaller ``sub-objects'' of scale $M' \ll M$. Indeed, let us consider a discrete model where a mass ${\cal M}$ (per unit volume) is recognized as ${\cal N}_i$ objects of mass $M_i=\lambda^i {\cal M}$ with $\lambda<1$ (e.g. $\lambda=1/2$) at any scale $i \geq 1$. In the logarithmic interval $dM/M = \Delta \ln M = \ln \lambda$ there are ${\cal N}_i = {\cal M}/M_i$ objects hence $\eta(M) \propto 1/M$ and $\theta=0$. Thus, the model described in the previous section can be made to satisfy ${\bf C1}$ and ${\bf C2}$ if we give the mass function $\eta(M) dM/M$ a new interpretation: it now counts the substructures of mass $M$ embedded within larger halos as well as new isolated objects. Moreover, the latter provide a negligible contribution to the moments $\mu_p$ ($p \geq 2$) as compared to the former. Hence, the previous considerations lead to a new simple and natural model:\\

${\bf H2}$ : {\it dark matter halos can be divided into substructures of arbitrarily small mass $M$ with a characteristic density equal to the mean density of the universe at the time when this mass scale became non-linear. Moreover, in this case the exponents $\gam_p$ ($p \geq 2$) are given by the stable-clustering approximation} (\ref{gampsc}).\\

Of course, logarithmic factors may be present: the mass embedded within substructures of scale lower than $M$ may slowly decrease as $1/|\ln M|$ for instance. One must note that ${\bf H2}$ does not necessarily contradict the result from numerical simulations that the {\it spherically averaged} density profile of dark matter halos follows a universal shape. Indeed, in such measures one automatically discards substructures in order to obtain a ``mean'' behaviour. However, it is surprising (and somewhat suspicious in our view) that this ``average profile'' does not depend on the slope $n$ of the initial power-spectrum while (at least) the first few correlation functions ($p \leq 6$) do depend on $n$ (indeed, as emphasized by Navarro et al.1997 although the characteristic density $\delta_c$ of $M_*$ halos decreases with $n$ they find that dark matter halos obey the same profile (\ref{rhot}) for all power-spectra). In particular, one would rather expect a density profile of the form $\rho(r) \propto r^{-\gam}$. We can also note that there is still a significant uncertainty on the shape of these halos (Moore et al.1998). The description ${\bf H2}$ predicts a significant amount of substructure (even with a logarithmic correction) while simulations often only have a moderate amount of ``sub-objects''. However, this overmerging declines with higher resolution so that it may be mostly a numerical problem due to low resolution (Ghigna et al.1998). In addition, as we noticed earlier the measure of the moments $\mu_p$ is a better defined operation than the identification of halos and substructures so that the conditions ${\bf C1}$ and ${\bf C2}$ are probably more robust than the results obtained for the characteristics of individual objects. Hence we think it is safer to rely on these two constraints than on observations about the behaviour of halos themselves. Thus, ${\bf H2}$ appears to be the simplest description which is consistent with numerical results, while ${\bf H1}$ is ruled out. We note that a simple model which leads to ${\bf H2}$ was described in Valageas (1998) (this article also described the quasi-linear regime for the density and velocity fields). A similar picture of the density field was also obtained by Balian \& Schaeffer (1989a) starting directly from the assumption that all exponents $\gam_p$ ($p \in \naturs$) satisfy the stable-clustering prediction (\ref{gampsc}) (see also Valageas \& Schaeffer 1997). We must stress that although the most simple interpretation of ${\bf H2}$ is that small clumps embedded within larger halos do not evolve much and keep roughly their initial density it can also describe a much more dynamical picture. Indeed, individual clumps may continuously form and disappear through the exchange of particles while the statistical properties of the density field remain unchanged.

\section{Multifractals}
\label{Multifractals}

\subsection{General model}

We have seen in Sect.\ref{Single power-law profile} and Sect.\ref{Mass-dependent profile} that one cannot describe the non-linear density field by a collection of smooth halos. Hence, we must turn to another description. The importance of the scaling exponents $\gam_p$ and the conditions ${\bf C1}$ and ${\bf C2}$ obviously lead one to consider multifractal models (e.g. Balian \& Schaeffer 1989b; Frisch 1995) to describe the non-linear regime. We first introduce the cumulative probability $\Pa_l(\rhot) = \int_{\rhot}^{2 \rhot} P_l(\rhot) d\rhot$ (for instance). Then, for any scaling exponent $\alpha \in \real$ we define the scaling dimension $F(\alpha) \leq 3$ (which can be negative and take the value $-\infty$) by:
\beq
\lim_{l \rightarrow 0} \; \frac{\ln \Pa_l(l^{\alpha})}{\ln l} = 3 - F(\alpha)
\label{Falpha}
\eeq
In particular, we assume that:\\

\begin{figure}[htb]

\centerline{\epsfxsize=8 cm \epsfysize=5.5 cm \epsfbox{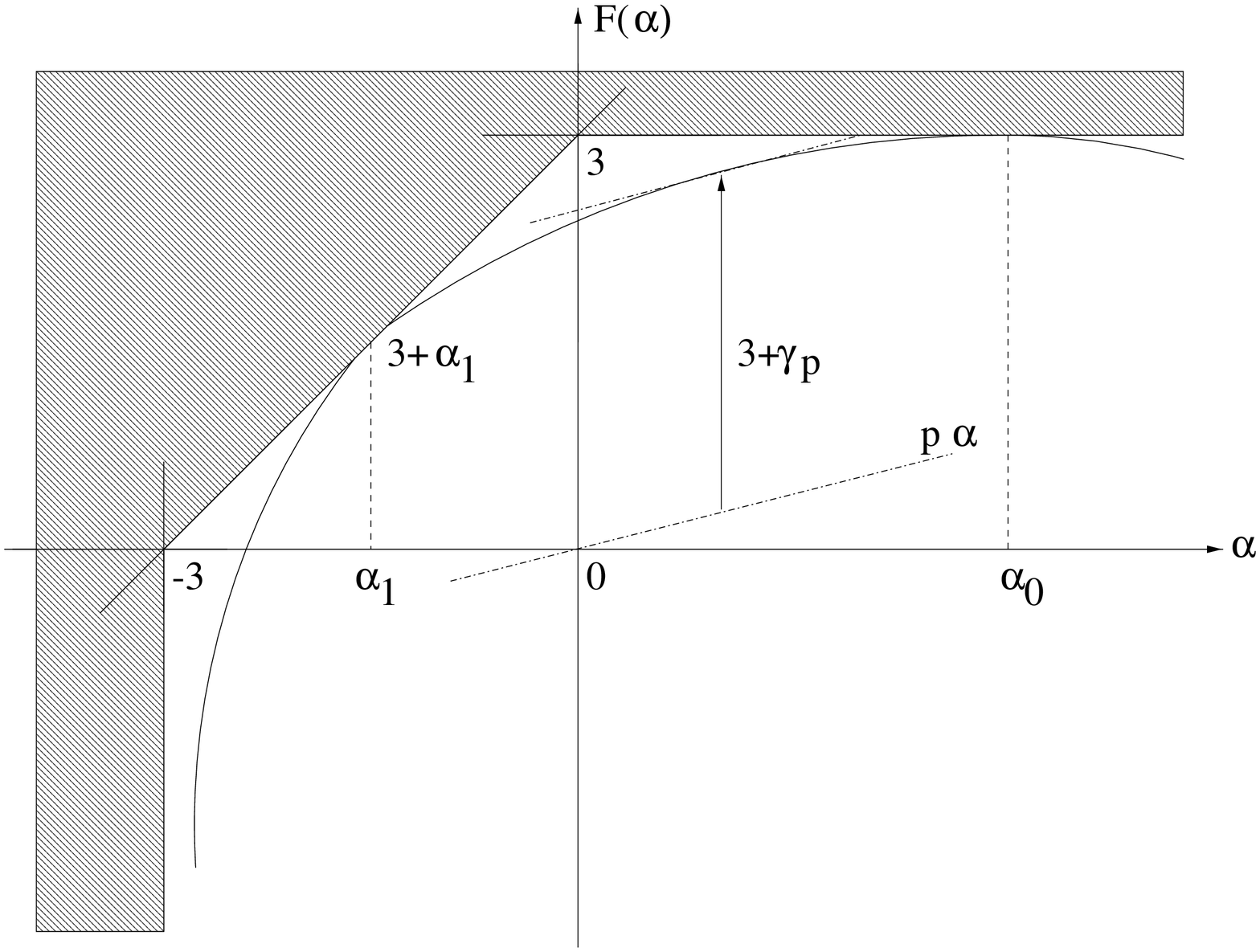}}

\caption{Geometrical construction of the scaling exponents $\gam_p$ from the fractal dimensions $F(\alpha)$.}
\label{fig1}

\end{figure}

{\bf H3} : {\it the density field is multifractal on small non-linear scales $l \ll l_*$ where:}
\beq
l \ll l_* \; : \hspace{0.5cm}  \Pa_l(l^{\alpha}) \sim l^{3 - F(\alpha)}
\eeq

where $l_*$ defined by $\sigma(l_*)=1$ corresponds to the transition between the linear and non-linear regimes. As usual we did not write possible logarithmic factors. Note that this definition of multifractality is less restrictive than the (more traditional) assumption: $\rhot_l({\bf r}) \sim l^{\alpha}$ for ${\bf r} \in {\cal D}_{\alpha} \subset \real^3$ with dim${\cal D}_{\alpha} = F(\alpha)$, for $l \rightarrow 0$. The scaling exponents $\gam_p$ defined by (\ref{gampl}) are obtained from ${\bf H3}$. Indeed:
\beq
\mu_p(l) = \int \rhot^{\;p} P_l(\rhot) d\rhot \sim  \int l^{p \alpha + 3 -F(\alpha)} \; P_*(\alpha) d\alpha
\eeq
where $P_*(\alpha)$ gives the weight of the various exponents. Then, using the steepest descent method, we obtain:
\beq
\gam_p = - \min_{\alpha} \left[ p \alpha + 3 -F(\alpha) \right] = \max_{\alpha} \left[ F(\alpha) - 3 - p \alpha \right]
\label{gampF}
\eeq
Thus, $\gam_p$ (seen as a function of $p \in \real$) and $F(\alpha)$ are related by a Legendre transformation. We note $\alpha_p$ the scaling exponent which corresponds to $\gam_p$. Then, we have from (\ref{gampF}):
\beq
p' > p \; : \hspace{0.5cm} \alpha_{p'} \leq \alpha_p
\label{seralp}
\eeq
We also get:
\beq
p \in \real \; : \hspace{0.5cm} F'(\alpha_p) = p \hspace{0.5cm} \mbox{if } F'(\alpha_p) \mbox{ exists}
\label{Fprime}
\eeq
Using (\ref{gamp3}) we obtain:
\beq
\alpha \geq -3
\label{alpha3}
\eeq
In other words $F(\alpha) = - \infty$ for $\alpha<-3$. Since $\gam_0=0$ there exists at least one point $\alpha_0$ such that $F(\alpha_0)=3$. Using $\gam_1=0$ we also obtain: $F(\alpha) \leq 3+\alpha$ and this equality is realized at least for one point $\alpha_1$. Thus we have:
\beq
F(\alpha_0)=3 \hspace{0.3cm} , \hspace{0.3cm} F(\alpha_1)=3+\alpha_1 \hspace{0.3cm} , \hspace{0.3cm} \alpha_1 \leq 0 \leq \alpha_0
\eeq
We show in Fig.\ref{fig1} an example of $F(\alpha)$ and how one can obtain the exponents $\gam_p$ from this curve.

The results of Sect.\ref{Single power-law profile}, where the density field was modelled as a collection of halos with a single power-law density profile, can be obtained from this multifractal description. Indeed, using the usual definition, in this case there are only two fractal dimensions: $\alpha=0$ with $F(0)=3$ (which corresponds to points in the halo of the objects or within a uniform component) and $\alpha=-\beta$ with $F(-\beta)=0$ (centers of the halos). Here we considered a finite number of halos since small objects do not contribute to $\mu_p$ for $p \geq 1$. Then one recovers (\ref{betagamp}) from (\ref{gampF}). Note that in this case $\alpha_0 = \alpha_1 = 0$. We have already seen in Sect.\ref{Single power-law profile} that such a model is ruled out.

\subsection{Bifractal}

The simplest model which is consistent with ${\bf C1}$ and ${\bf C2}$ is the case of a bifractal, where there are only two scaling exponents: $\alpha = \alpha_1<0$ and $\alpha = \alpha_0 > 0$. Then, ${\bf C1}$ implies $\alpha_1=-\gam$ and we note $\alpha_0 = \gam \om /(1-\om)$ with $0<\om<1$. Thus:
\beq
\left\{ \begin{array}{l} {\displaystyle F(-\gam) = 3-\gam \hspace{0.5cm} , \hspace{0.5cm} F\left(\frac{\gam \om}{1-\om} \right) = 3 } \\ \\  F(\alpha) = -\infty \hspace{0.5cm} \mbox{otherwise} \end{array} \right.
\label{bif1}
\eeq
and:
\beq
\left\{ \begin{array}{lcl} p > 1-\om & : & \gam_p = (p-1) \; \gam  \\ \\ p < 1-\om & : & \gam_p = - p \; \frac{\gam \om}{1-\om} \end{array} \right.
\label{bif}
\eeq
This bifractal model is consistent with the stable-clustering assumption (\ref{scal1}), see (\ref{gampsc}). Moreover, it saturates the inequalities (\ref{zetap2}). We note that, starting from the behaviour (\ref{scal1}) for the many-body correlation functions, Balian \& Schaeffer (1989b) obtained the fractal dimensions implied by simple models for the density field which are consistent with these scaling laws and they recovered this bifractal model. More precisely, they found $F(\alpha) = -\infty$ for $\alpha<-\gam$ and $\alpha> \gam \om /(1-\om)$, while in the intermediate range $-\gam < \alpha < \gam \om /(1-\om)$ the dimension $F(\alpha)$ must be lower than (or equal to) the line joining $(-\gam,3-\gam)$ to $(\gam \om /(1-\om),3)$. Thus, these points play no role for the exponents $\gam_p$ which are still given by (\ref{bif}). 

On the other hand, we can note that a larger class of models (than the bifractal) is consistent with the stable-clustering prediction (\ref{gampsc}) for the correlation functions since one only needs:
\beq 
F(\alpha)=-\infty \hspace{0.2  cm} \mbox{for} \hspace{0.2cm} \alpha<-\gam \hspace{0.5cm} \mbox{and} \hspace{0.5cm} F(-\gam)=3-\gam
\eeq
The exponents $\gam_p$ for small $p$ are related to the behaviour for large $y$ of the function $\varphi_l(y)$ introduced in (\ref{phiy}). Thus, if we have:
\beq
y \rightarrow \infty \; : \hspace{0.3cm} \varphi(y) \sim a \; y^{1-\om} \hspace{0.3cm} \mbox{with} \hspace{0.3cm}  0 < \om < 1
\label{phiom1}
\eeq 
where $\varphi(y) = \lim_{l \rightarrow 0} \varphi_l(y)$ we recover the second line of the system (\ref{bif}), as shown in appendix in (\ref{muom}). On the other hand, if $\varphi(y) \sim \ln^2 y$ for $y \rightarrow \infty$ one obtains $\gam_p = \infty$ for $p<0$. This corresponds to density fields where there is a large probability to find very underdense regions while (\ref{phiom1}) leads to an exponential cutoff for $\rhot \rightarrow 0$, see (\ref{Pz}). 

We note that this model is consistent with a universal spherically averaged density profile for virialized halos of the form described in Sect.\ref{Single power-law profile} or Sect.\ref{Mass-dependent profile}. It simply adds to the two scaling exponents $-\gam$ and $\gam \om /(1-\om)$ the points $(-\beta,0)$ (for Sect.\ref{Single power-law profile} with now $\beta \leq \gam$) or $(-\gam,3-\gam-6 \theta/(5+n))$ (for Sect.\ref{Mass-dependent profile}, but in this case this point does not show on the curve $F(\alpha)$ since $\theta>0$). However, even if such a ``universal shape'' exists, it does not govern the exponents $\gam_p$ which are given by the internal density fluctuations whose contribution dominates the moments $\mu_p$ and the relations (\ref{bif}) are still valid. We also note that the scaling dimension $F(-\gam)=3-\gam$ which characterizes the matter distribution can lead to the filaments and walls seen in numerical simulations (e.g. Bond et al.1996) since $1 \leq 3-\gam \leq 2$.

\subsection{Specific constraints}

In order to obtain more detailed characteristics of the curve $F(\alpha)$ we must explicitely consider the specific case we are interested in: the growth through gravitational instability of small gaussian density fluctuations, in a critical universe. Although one should obtain the exponents $\gam_p$ and $F(\alpha)$ from the Liouville equation, which is not an easy task, it is possible to derive a few constraints from simple and reasonable arguments.

We first consider the behaviour of underdense regions, which governs the exponents $\gam_p$ for low $p$ and $F(\alpha)$ for large $\alpha$. We assume that strong underdensities (``voids'') are created by the amplification of small initial negative density fluctuations. In particular, at a given scale $R$ very low density regions ($(1+\delta) \rightarrow 0$) come from the expansion of rare initial underdensities. In the early linear universe, density fluctuations at the mass scale $M$ are characterized by the parameter $\nu$:
\beq
\nu = \frac{\delta_L}{\sigma(M)}
\eeq
where $\delta_L$ is the linear density contrast. As seen in Valageas (1998) rare underdensities $\nu \ll -1$ have a spherically symmetric profile. Moreover they tend to become increasingly spherical as they expand (Bertschinger 1985), hence we can use the spherical model to obtain their dynamics (see also Bernardeau 1994b) before their expansion is slowed down when they join with neighbouring underdensities. Thus, we write that the probability to find an underdensity $(1+\delta) \rightarrow 0$ in a cell of radius $R$ is smaller than the probability we would obtain if all initial underdensities would follow the spherical dynamics expansion. Thus, we write:
\beq
P_R(<1+\Delta) \leq \int_{-\infty}^{\delta_L(\Delta)} \frac{1}{(1+\delta)(\delta_L)} \; P_L(\delta_L) d\delta_L
\eeq
where $P_L(\delta_L) d\delta_L$ is the probability distribution of the initial linear density fluctuations:
\beq
P_L(\delta_L) = \frac{1}{\sqrt{2\pi}\sigma} \; e^{-\delta_L^2/(2 \sigma^2(R_m))}
\eeq
Here, $R_m$ plays the role of a mass scale so that $R_m = (1+\Delta)^{1/3} R$. In the limit $(1+\delta) \rightarrow 0$, the spherical model relates the actual density contrast to the linear theory prediction by (Peebles 1980; Valageas 1998):
\beq
\delta_L \rightarrow -\infty \; : \hspace{0.5cm} (1+\delta) \sim \left( - \frac{20}{27} \; \delta_L \right)^{-3/2}
\eeq
Thus, we obtain:
\beq
\begin{array}{l} {\displaystyle P_R(<1+\delta) \leq \int_{-\infty}^{-27/20 \; (1+\delta)^{-2/3}} \frac{d \delta_L}{\sqrt{2\pi}\sigma} \;  \left( - \frac{20}{27} \; \delta_L \right)^{3/2} } \\ \\ {\displaystyle \hspace{3cm} \times \; \exp \left[ - \frac{\delta_L^2}{2 \sigma^2(R_m)} \right] } \end{array}
\eeq
which leads to:
\beq
\begin{array}{l}  {\displaystyle \mbox{for} \hspace{0.3cm} \rhot^{\;-\frac{1-n}{3}} \left( \frac{l}{l_*} \right)^{n+3} \gg 1 \; : } \\ \\ {\displaystyle P_l(<\rhot) \leq \frac{25}{27 \sqrt{2\pi}} \;  \rhot^{\;-\frac{n+5}{6}} \left( \frac{l}{l_*} \right)^{-\frac{n+3}{2}} }  \\ \\ {\displaystyle  \hspace{1.5cm} \times \; \exp \left[ - \frac{1}{2} \left(\frac{27}{20}\right)^2 \rhot^{\;-\frac{1-n}{3}} \left( \frac{l}{l_*} \right)^{n+3} \right] } \end{array}
\label{Pspher}
\eeq
This means that $\lim_{l \rightarrow 0} \ln P_l(<l^{\alpha}) / \ln l = \infty$ in the case $-\alpha (1-n)/3 + (n+3) <0$. From the definition (\ref{Falpha}) we get:
\beq
F(\alpha) = - \infty \hspace{0.5cm} \mbox{for} \hspace{0.5cm} \alpha > \frac{3(3+n)}{1-n}
\label{alphasup}
\eeq
In the framework of the bifractal model (\ref{bif1}) it implies:
\beq
\mbox{fractal} \; : \hspace{0.5cm} \om \leq \frac{5+n}{6}
\label{omspher1}
\eeq
However, it is possible to obtain another upper limit for $\om_l$ without assuming the density field to be a multifractal. Indeed, let us assume as in (\ref{phiom}) that $\varphi_l(y)$ is a power-law for large $y$ with an exponent $\om_l$ which may depend on the scale $l$ (the bifractal model is the peculiar case where $\om_l$ and $a_l$ are scale-invariant). Then the low-density cutoff of the density probability distribution $P_l(\rhot)$ is given by (\ref{Pz}). The constraint (\ref{Pspher}) applied in the limit $\rhot \rightarrow 0$ implies $(1-\om_l)/\om_l \geq (1-n)/3$ hence we obtain the scale-independent upper bound:
\beq
l \ll l_* \; : \hspace{0.5cm} \om_l \leq \frac{3}{4-n}
\label{omspher2}
\eeq
Here $\om_l$ is not necessarily related to a multifractal dimension, it only characterizes the density probability distribution seen at scale $l$. The existence of this upper bound is due to the fact that the probability to have very low densities decreases strongly as $\rhot \rightarrow 0$. Indeed, in the linear regime the density field is very close to uniform (because of the sharp cutoff of the gaussian) and the amplification of underdensities is only a power-law of time in both the linear and non-linear regimes (but with different exponents). Hence at late times the density field still ``remembers'' the initial gaussian cutoff. Our result is only based on the assumption that the most extreme ``voids'' come from the amplification of the most negative primordial density fluctuations which seems quite reasonable. We did not take into account shell-crossing since this would rather affect collapsed regions (it does not occur for the spherically symmetric voids considered here, see Valageas 1998). Note that for the values of $n$ of interest the bound (\ref{omspher2}) is more constraining than (\ref{omspher1}). We compare in Tab.1 the values of $\om$ obtained in numerical simulations with the upper bounds (\ref{omspher1}) and (\ref{omspher2}).

\begin{table}
\begin{center}
\caption{Exponent $\omega$ for various indexes $n$ of the power-spectrum. We compare the upper bounds (\ref{omspher1}) and (\ref{omspher2}) with numerical results from Colombi et al.(1997) (line C), Munshi et al.(1999) (line M) and Valageas et al.(1999a) (line V).}

\begin{tabular}{ccccc}\hline

$n$ & $-2$ & $-1$ & $0$ & $1$ \\ 
\hline\hline
\\ 

$(5+n)/6$ & 0.5 & 0.66 & 0.83 & 1  \\

$3/(4-n)$ & 0.5 & 0.6 & 0.75 & 1  \\
\hline\hline
\\

C & 0.3 & 0.5 & 0.65 & 0.65 \\

M & 0.33 & 0.4 & 0.55 & 0.7 \\

V & 0.3 & 0.4 & 0.45 &

\end{tabular}
\end{center}
\end{table}

We can see that the numerical results are consistent with the upper bounds (\ref{omspher1}) and (\ref{omspher2}). More generally, although our argument was not completely rigorous the results (\ref{alphasup}) and (\ref{omspher2}) should be quite robust. Indeed, one expects deviations from spherical dynamics to slow down the decrease of the ``overdensity'' $(1+\delta)$ which leads to smaller $\alpha$ and $\om$. Indeed, a lower $\alpha$ or $\om$ means less extreme underdensities. This may also explain why the values of $\om$ obtained from numerical simulations are smaller than the upper bound (\ref{omspher2}).

Now, we show that the existence (for $n<1$) of an upper limit smaller than unity for $\om_l$ implies that $\om_l$ is {\it scale-independent} within the framework of a multifractal model. Since $\alpha$ admits a finite upper bound the sequence $\alpha_{-p}$ has a finite limit $\alpha_+$ for $p \rightarrow +\infty$, see (\ref{seralp}). Moreover, we have from (\ref{gampF}):
\beq
\begin{array}{ll} {\displaystyle  \gam_{-(p+1)} - \gam_{-p} = } & {\displaystyle \alpha_{-(p+1)} \; + \; p \left( \alpha_{-(p+1)} -  \alpha_{-p} \right) } \\  \\  & {\displaystyle + \; F \left( \alpha_{-(p+1)} \right) -  F \left( \alpha_{-p} \right) } \end{array}
\eeq
If $F(\alpha_+)$ is finite and $\alpha_+$ is isolated, or $F'(\alpha_+)$ is finite, or $F'(\alpha_+)=+\infty$, there exists a real $p_+$ such that for any $p \geq p_+$ we have $\alpha_{-p} = \alpha_+$. If $F(\alpha)$ is twice differentiable on an interval $[\hat{\alpha},\alpha_+[$ with $\hat{\alpha} < \alpha_+$ and $F'(\alpha_+) = -\infty$ then $\alpha_{-p} < \alpha_+$ for any finite $p$ and using a Taylor expansion with remainder and (\ref{Fprime}) we obtain $( \gam_{-(p+1)} - \gam_{-p}) = \alpha_{-(p+1)} + R_p$ with $|R_p| \leq ( \alpha_{-(p+1)} -  \alpha_{-p} )$. Thus, in all these cases we get:
\beq
\lim_{p \rightarrow +\infty} \left( \gam_{-(p+1)} - \gam_{-p} \right) = \alpha_+ \hspace{0.3cm} \mbox{which is finite}
\label{alphap}
\eeq
Using (\ref{muom}) we have:
\beq
p \rightarrow +\infty \; : \hspace{0.5cm} \ln \left[ \frac{\mu_{-(p+1)}(l)}{\mu_{-p}(l)} \right] \sim \frac{\om_l}{1-\om_l} \; \ln p
\eeq
From the definition (\ref{gampl}) of the exponents $\gam_p$ we also have:
\beq
\begin{array}{l} {\displaystyle l \ll l_* \; , \; p \in \real \; : \hspace{0.3cm}  \ln \left[ \frac{\mu_{-(p+1)}(l)}{\mu_{-p}(l)} \right] = \ln \left[ \frac{\mu_{-(p+1)}(l_*)}{\mu_{-p}(l_*)} \right] } \\ \\ {\displaystyle \hspace{3.7cm}  - \left( \gam_{-(p+1)} - \gam_{-p} \right) \ln \left( \frac{l}{l_*} \right) } \end{array}
\eeq
Here $\ln(\mu_p(l_*))$ is the value obtained in $l=l_*$ by extrapolation from the non-linear regime $l \rightarrow 0$. Thus, we obtain:
\beq
\begin{array}{l} {\displaystyle l \ll l_* \; : \hspace{0.5cm} \frac{\om_l}{1-\om_l} = \frac{\om_{l_*}}{1-\om_{l_*}} } \\ \\ {\displaystyle \hspace{3cm} - \ln \left( \frac{l}{l_*} \right) \; \lim_{p \rightarrow \infty} \frac{\gam_{-(p+1)} - \gam_{-p}}{\ln p} } \end{array}
\label{omlct}
\eeq
Here again $\om_{l_*}$ is defined by the asymptotic non-linear regime. Using (\ref{alphap}) we see that $\om_l$ is constant, apart from a possible sub-logarithmic $l-$dependence (e.g. $\ln[\ln(1/l)]$) which can be safely neglected for any practical purposes. Thus, the assumption of a multifractal together with the existence (for $n<1$) of an upper bound for the scaling exponents $\alpha$ is sufficient to show that the exponent $\om$ of the power-law regime of $\varphi_l(y)$ at large $y$ is scale-independent. However, the normalization factor $a_l$ may vary with $l$. Indeed, from (\ref{muom}) and (\ref{gampl}) we obtain:
\beq
\frac{a_l}{a_{l_*}}  = \left( \frac{l}{l_*} \right)^{(1-\om) (\alpha_+ - \alpha_0)} \propto \xia^{\; - (1-\om) (\alpha_+ - \alpha_0) /\gam}
\label{alap}
\eeq
Thus, the measure of the normalization factor $a_l$ allows one to get an important constraint on the curve $F(\alpha)$: it provides the value of the maximum scaling exponent $\alpha_+$. Indeed, we have:
\beq
\alpha > \alpha_+ \; : \hspace{0.5cm} F(\alpha)=-\infty
\eeq
Note that $\alpha_+ \geq \alpha_0$ and $\alpha_+$ is lower than the upper bound (\ref{alphasup}) while the upper limit (\ref{omspher2}) is irrelevant. Thus, the coefficient $a_l$ remains constant (neglecting logarithmic corrections) or tends to 0 as $\xia \rightarrow \infty$ (which means that underdense regions have an even smaller density). We note that numerical simulations (Colombi et al.1997; Munshi et al.1999; Valageas et al.1999a) for the counts-in-cells are consistent with a constant value of $a_l$ while the behaviour of the mass functions studied in Valageas et al.(1999a) might suggest a very slow decrease of $a_l$ at smaller scales. From the dispersion of the measure of $a_l$ in these numerical results and the small decline of the mass functions we can obtain an upper bound for $\alpha_+$ using (\ref{alap}). We show our results in Tab.2. (to get $\alpha_+$ we used the mean $\om$ of the various simulations in (\ref{alap})).

\begin{table}
\begin{center}
\caption{Exponents $\alpha$ for various indexes $n$ of the power-spectrum. We compare the upper bound (\ref{alphasup}) with the mean value $\alpha_0$ given by numerical results from Colombi et al.(1997), Munshi et al.(1999) and Valageas et al.(1999a). We also present the maximum exponent $\alpha_+$ allowed by simulations, using (\ref{alap}). The exponent $\gam$ is given by ${\bf C1}$ in agreement with numerical results. The upper bound for $\alpha_-$ is obtained from the measured dispersion of $y_s$ using (\ref{ysa}).}

\begin{tabular}{ccccc}\hline

$n$ & $-2$ & $-1$ & $0$ & $1$ \\ 
\hline\hline
\\ 

$3(3+n)/(1-n)$ & 1 & 3 & 9 & $\infty$  \\

$\alpha_0$ & 0.45 & 1.13 & 2.2 & 4.15  \\

$\alpha_+$ & 0.59 & 1.39 & 2.6 & 4.42  \\
\hline\hline
\\

$\gam$ & 1 & 1.5 & 1.8 & 2 \\

$\alpha_-$ & 1.15 & 1.70 & 2.07 & 2.19

\end{tabular}
\end{center}
\end{table}

Thus, the behaviour of $F(\alpha)$ in the domain $\alpha>\alpha_0$ is very close to the bifractal model. Indeed, numerical simulations show that $F(\alpha)=-\infty$ for scaling exponents $\alpha$ very close to $\alpha_0$. Moreover, even if there exist scaling exponents slightly larger than $\alpha_0$ (which may not be the case) they do not affect the slope $\om$ of the density probability distribution while the normalization factor $a_l$ only shows a very weak (if any) scale dependence. We note that the value of $\alpha_+$ obtained from numerical simulations through (\ref{alap}) is significantly lower than the theoretical upper bound (\ref{alphasup}) and very close to $\alpha_0$. This suggests that $\alpha_+$ may in fact be equal to $\alpha_0$. Indeed, if one measures directly the curve $F(\alpha)$ in a numerical simulation, using (\ref{Falpha}) for instance, one would not be able to obtain $F(\alpha_0)=3$ and $F(\alpha)=-\infty$ for all larger scaling exponents. Due to finite resolution effects there would be some dispersion which would create some spurious exponents slightly larger than $\alpha_0$. Moreover, it is obvious from (\ref{Falpha}) that the sharp $F(\alpha)$ which appears for instance in the bifractal model is only realized in the limit $l \rightarrow 0$: for any simulation where the available range of $l$ is necessarily finite the ratio $-\ln \Pa_l(l_{min}^{\alpha}) / \ln l_{min}$ will show a smooth cutoff.

We now turn to high density regions (i.e. small $\alpha$). Let us assume that after collapse mass condensations see their overdensity evolve as: $(1+\delta) \propto a^{3 \beta}$. If these halos keep the same density (or radius) with time (i.e. no evolution) while they get embedded within larger objects we have $\beta=1$. Since $\delta_L \propto a$ (and collapse occurs for $\delta_L \sim 1$) we obtain for high densities: $(1+\delta) \sim \delta_L^{3\beta}$. Hence we get:
\beq
\nu \sim (1+\delta)^{\frac{1}{3\beta}+\frac{n+3}{6}} \; R^{\frac{n+3}{2}}
\eeq
and:
\beq
F(\alpha) = - \infty \hspace{0.5cm} \mbox{for} \hspace{0.5cm} \alpha < - \frac{3(3+n)}{n+3+2/\beta}
\label{alphasup2}
\eeq
Using the constraint ${\bf C1}$ we get $\beta \geq 1$. Thus, there are two possibilities within this description: i) after virialization collapsed halos do not see their properties (density, radius) evolve much (i.e. at most through logarithmic factors) and the exponents $\gam_p$ ($p \geq 2$) are given by the stable-clustering assumption (\ref{gampsc}) or ii) some objects see their density {\it increase} with time as a power-law and the exponents $\gam_p$ ($p \geq 2$) are larger than the values (\ref{gampsc}). As larger scales become non-linear and small objects get embedded within increasingly massive halos, interactions between clouds (e.g. collisions) can affect their properties. Although one might have expected these processes to decrease the mean density of small objects (through disruptions, relaxation towards the lower density of the larger host halo) the constraints ${\bf C1}$ and ${\bf C2}$ show on the contrary that their mean density is constant (except for a possible logarithmic decline) or increases. This description assumes that one can follow the behaviour of individual mass condensations. In fact, this may not be the case as density fluctuations could be continuously created and destroyed, so that particles belong alternatively to overdensities and underdensities of evolving magnitude. Thus, the behaviour of mass condensations could be more intricate than those of ``voids''. In particular, one should explicitely take into account shell-crossing. However, we can directly obtain constraints on the curve $F(\alpha)$ from numerical results. The location $y_{s,l} \leq 0$ of the singularity of $\varphi_l(y)$, see (\ref{ys}) and (\ref{phiy}), is given by:
\beq
y_{s,l} = - \lim_{p \rightarrow \infty} \frac{p \; S_p(l)}{S_{p+1}(l)}
\label{ysSp}
\eeq
As explained in appendix we assume here that $y_{s,l}$ is finite and non-zero. Then, we obtain from (\ref{zetap}):
\beq
y_s(l) = y_s(l_*) \; \lim_{p \rightarrow \infty} \left( \frac{l}{l_*} \right)^{\gam_{p+1}-\gam_p-\gam}
\eeq
where again $y_s(l_*)$ is defined from the non-linear regime. Since $\alpha$ admits a finite lower bound (\ref{alpha3}) (only due to the positivity and additivity of the mass) the sequence $\alpha_{p}$ has a finite limit $-\alpha_-$ for $p \rightarrow +\infty$, see (\ref{seralp}). From (\ref{gampF}) we obtain:
\beq
\lim_{p \rightarrow +\infty} \left( \gam_{p+1} - \gam_{p} \right) = \alpha_- \hspace{0.3cm} , \hspace{0.3cm} \gam \leq \alpha_- \leq 3
\label{alpham}
\eeq
in a fashion similar to (\ref{alphap}). Hence we have:
\beq
y_s(l) = y_s(l_*) \; \left( \frac{l}{l_*} \right)^{\alpha_- -\gam}
\label{ysa}
\eeq
Thus, if $P_l(\rhot)$ shows a pure exponential cutoff at some scale $l$ (i.e. $y_s(l)$ is finite and non-zero) it will display a similar exponential cutoff throughout the non-linear regime. Moreover, the strength $y_s(l)$ of this falloff declines or remains constant at smaller scales: the importance of extreme positive density fluctuations grows or is stationary as one probes deeper into the non-linear regime (as compared with the predictions of the stable-clustering ansatz). From the dispersion of the measure of $y_{s,l}$ obtained in numerical simulations (Colombi et al.1997; Munshi et al.1999; Valageas et al.1999a) we obtain using (\ref{ysa}) an upper bound for $\alpha_-$. We show our results in Tab.2. Note that we have:
\beq
\alpha < - \alpha_- \; : \hspace{0.5cm} F(\alpha)=-\infty
\eeq
Thus, as was the case for the large $\alpha$ domain discussed above numerical simulations constrain the curve $F(\alpha)$ to be rather close to the bifractal model for $\alpha \leq -\gam$. The possible scaling exponents do not extend down to $\alpha=-3$ and their minimum value $-\alpha_-$ is close to $-\gam$. This suggests again that the existence of scaling exponents below $-\gam$ may be due to finite effects (note moreover that the values displayed in Tab.2 are upper bounds). Although we assumed $y_{s,l}$ to be finite and non-zero the constraint on $\alpha_-$ shown in Tab.2 also applies if the cutoff is not a ``pure'' exponential. Thus, if we write more generally:
\beq
\rhot \gg \xia \; : \; P_l(\rhot) \simeq  \frac{a_{s,l}}{\xia^{\;2}} \left( \frac{\rhot}{\xia} \right)^{\omega_{s,l}-1}  \exp \left[ - \left( \frac{\rhot}{x_{s,l} \; \xia} \right)^{\kappa_l} \right]  
\eeq
with $\kappa_l > 0$ we obtain:
\beq
p \rightarrow \infty \; : \; \mu_p \sim \xia^{\;p-1} \; \frac{a_{s,l}}{\kappa_l} \; x_{s,l}^{p+\om_{s,l}} \; \Gamma \left( \frac{p+\om_{s,l}}{\kappa_l} \right)
\label{mupkap}
\eeq
The case $\kappa_l>1$ corresponds to $y_{s,l}=-\infty$ (no singularity) and $\kappa_l<1$ to $y_{s,l}=0$. In a fashion similar to (\ref{omlct}), using the constraint $\alpha \geq -3$, we obtain that $\kappa_l$ is constant (except for sub-logarithmic terms). From (\ref{mupkap}) we also obtain:
\beq
x_s(l) = \frac{1}{\;\xia\;} \; \lim_{p \rightarrow \infty} \left( \frac{\kappa_l}{p} \right)^{1/\kappa_l} \; \frac{\mu_{p+1}}{\mu_p}
\eeq
which is similar to (\ref{ysSp}). Since $\kappa_l$ exhibits at most a sub-logarithmic dependence this implies:
\beq
x_{s}(l) = x_{s}(l_*) \; \left( \frac{l}{l_*} \right)^{-(\alpha_- -\gam)}
\label{xsa}
\eeq
which is exactly (\ref{ysa}) we got in the peculiar case $\kappa=1$ since $x_{s,l}=-1/y_{s,l}$. Note that (\ref{xsa}) does not depend on $\kappa$. Moreover, numerical results constrain the argument of the exponential to be close to a linear function of the density: $\kappa \simeq 1$.

Thus, we have shown that the curve $F(\alpha)$ is very similar to the bifractal model for $\alpha \leq \alpha_1$ and $\alpha \geq \alpha_0$. The intermediate regime is more difficult to constrain since it depends on the details of the dynamics. For instance, one could add a point $(0,3)$ if a finite fraction of the volume is filled by a smooth component. However, such modifications only influence the exponents $\gam_p$ for $0<p<1$, which corresponds to intermediate density fluctuations between the characteristic density of mass condensations (given by $x_{s,l} \xia$) and ``voids'' ($\rhot \sim a_l^{1/(1-\om)} \;  \xia^{\;-\om/(1-\om)}$). One needs to solve the equations of the dynamics in order to get more precise information. However, it is clear from the previous results that the bifractal model provides a reasonable description of the density field. Moreover, it is the simplest realistic model.

We note that in the bifractal model the density probability distribution $P_l(\rhot)$ in the very low-density regime (\ref{Pz}) also defines its behaviour in the intermediate regime (\ref{hs}), and conversely, through the parameters $a$ and $\om$. This also holds for more general models where $\varphi_l(y) \simeq a_l \; y^{1-\om}$ for large $y$. Indeed, as we noticed in Sect.\ref{Scaling exponents} the coefficients $S_p$ (or $\mu_p$) for $p \in \natur$ are sufficient to define entirely $P_l(\rhot)$ although they are mainly sensitive to positive density fluctuations. However, the parameters $\om$ and $a$ depend in a somewhat intricate fashion on the moments $\mu_p$ or $S_p$ ($p \in \natur$) through the asymptotic behaviour of the function $\varphi_l(y)$ while they can very easily be obtained theoretically from the moments $\mu_p$ ($-p \in \natur$) through (\ref{muom}). On the other hand, the parameters $\kappa$, $\omega_s$ or $x_{s,l}$ are directly given by the moments $\mu_p$ or $S_p$ ($p \in \natur$) through (\ref{mupkap}). Thus, both sequences have their own advantages.

\section{Conclusion}

In this article we have investigated the consequences for the shape of virialized halos of the constraints provided by numerical simulations on the first few correlation functions. Thus, we have shown that the density field {\it cannot} be described by a collection of {\it smooth} halos with a universal density profile. This does not imply that a mean spherically averaged density profile does not exist. It means that substructures within larger objects play an important role. In particular, we have shown that a possible interpretation of the constraints provided by numerical simulations is that dark matter halos can be divided into an infinite hierarchy of smaller objects. Then, the characteristic density of substructures of a given mass is the density of the universe at the time when this mass scale turned non-linear. However, these small mass condensations may not be permanent entities as they could be continuously created and destroyed through the long-range action of gravity. Finally, we have presented multifractal models which can reproduce the observed behaviour of the first few correlation functions. Thus, it appears that the {\it simplest realistic model} of the non-linear density field is the {\it bifractal} model developped by Balian \& Schaeffer (1989a, 1989b). Moreover, we have shown that numerical results constrain the multifractal properties of the actual non-linear density field, and the probability distribution of the density contrast, to be very close (or identical) to the characteristics of such a bifractal. In the main text we have only considered the case of a critical universe with an initial power-spectrum which is a power-law. Nevertheless, our main results (in particular the importance of substructures and small-scale density fluctuations) should also apply to power-spectra which are not pure power-laws but smooth enough to be reasonably approximated by a power-law over some significant range of mass (e.g. CDM). However, in such a case the scaling exponents $\gam_p$ may exhibit a slow scale-dependence as a function of the local slope of the power-spectrum. Our results should also apply to a low-density universe. In fact, since in such a model non-linear scales collapsed when the universe was still close to critical the non-linear regime should be the same as for the case $\Omega=1$ with the same power-spectrum. One only needs to take care of the different time-dependence of the normalization of $\xia$ in the non-linear part when $\Omega$ gets small, as done in Peacock \& Dodds (1996) and Valageas \& Schaeffer (1997). The implications for astrophysical objects of this description have already been presented in other studies (e.g. Valageas \& Schaeffer 1998 for galaxies; Valageas et al.1999b for Lyman-$\alpha$ clouds).

\vspace{1cm}	

\appendix

{\bf APPENDIX}

\section{Density probability distribution}
\label{Density probability distribution}

Here we briefly recall the behaviour of the probability distribution of the overdensity $\rhot_l$ in spheres of size $l$ implied by simple forms of $\varphi_l(y)$ (see Balian \& Schaeffer 1989a for details). The relation (\ref{Pl}) can be inverted as:
\beq
e^{-\varphi_l(y)/\xia} = \int_{0}^{\infty} e^{-\rhot \; y/\xia} \;\; P_l(\rhot) d\rhot   
\label{phiP(rho)}
\eeq
This implies (using $\mu_1=1$):
\beq
\begin{array}{l} {\displaystyle  \mbox{Re}(y) \geq 0 \; : \; \mbox{Re} \left[ \varphi_l(y) \right] \geq 0 \hspace{0.3cm} , \hspace{0.3cm} y \in \real : \; \varphi_l'(y) > 0 } \\  \\  {\displaystyle  y \in \real^+ : \; \varphi_l(y) \leq \xia \ln 2 + 2 y } \end{array}
\label{consphi}
\eeq
We assume that $\varphi_l(y)$ behaves as a power-law for large $y$:
\beq
y \rightarrow \infty \; : \hspace{0.3cm} \varphi_l(y) \sim a_l \; y^{1-\om_l} \hspace{0.3cm} \mbox{with} \hspace{0.3cm}  0 \leq \om_l \leq 1
\label{phiom}
\eeq 
The bounds on $\om_l$ are due to (\ref{consphi}). In the following we only consider $0 < \om_l < 1$ which is consistent with numerical simulations (Colombi et al.1997; Munshi et al.1999; Valageas et al.1999a) which give $0.3 \leq \om_l \leq 0.7$ for $-2 \leq n \leq 1$. Note that since we do not assume the coefficients $S_p(l)$ to be scale-invariant we write explicitely the $l-$dependence of the function $\varphi_l(y)$ and its parameters $a_l$, $\om_l$. Then, we consider the case where the rapid growth of the coefficients $S_p$ with $p$ implies a singular behaviour of $\varphi_l(y)$ at small negative values of $y$, say $y_{s,l} = -1 / x_{s,l}$ with $x_{s,l}$ being large ($x_{s,l} \sim 10$):
\beq
y \rightarrow y_{s,l}^{+}  \; : \;\; \varphi_l(y) = - a_{s,l} \Gamma(\omega_{s,l}) \; (y-y_{s,l})^{-\omega_{s,l}}
\label{ys}
\eeq
where we neglected less singular terms. This leads to an exponential cutoff at large densities of $P_l(\rhot)$ which agrees with numerical simulations (Colombi et al.1997; Valageas et al.1999a) and this singular behaviour of $\varphi_l(y)$ arises naturally from a tree-model for the correlation functions (Bernardeau \& Schaeffer 1992).

\begin{figure}[htb]

\begin{picture}(230,180)(-18,-5)

\epsfxsize=8 cm
\epsfysize=14 cm
\put(-8,-118){\epsfbox{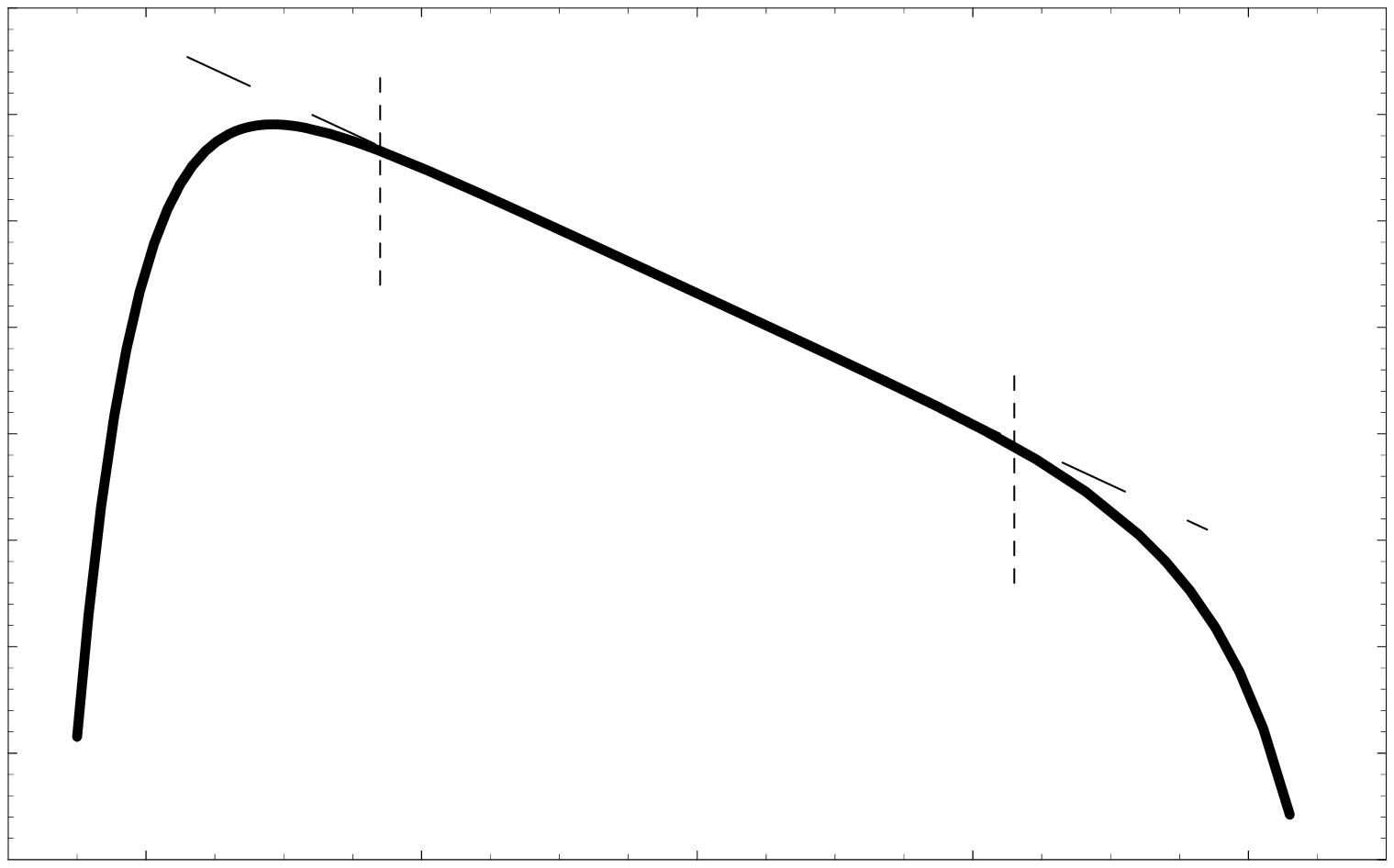}}
\put(25,-12){-4}
\put(111,-12){0}
\put(195,-12){4}
\put(-10,41){-10}
\put(-2,123){0}
\put(59,103){$\rhot_v$}
\put(158,45){$\rhot_c$}
\put(170,150){$\log[P_l(\rhot)]$}
\put(160,9){$\log(\rhot)$}
\put(60,35){$\xia = 200$}

\end{picture}

\caption{Probability $P_l(\rhot)$ for finding an overdensity $\rhot$ in a cell within which the average correlation function is $\xia=200$ in the case $\omega_l=1/2, \; \omega_{s,l}=-3/2$ and $x_{s,l}=10$. The density distribution is a power-law with a cutoff at small ($\rhot_v \sim \xia^{\; -\omega_l/(1-\omega_l)}$) and large ($\rhot_c \sim \xia$) densities. Both $\rhot_v$ and $\rhot_c$ are scale-dependent: for large $\xia$, $\rhot_v$ goes to 0 while $\rhot_c$ goes to infinity. The dashed line is the approximation (\ref{hs}): $P_l(\rhot) \propto \rhot^{\omega_l -2}$.}
\label{figPrho}

\end{figure}

This form of $\varphi_l(y)$ leads to a specific behaviour for $P_l(\rhot)$ in the non-linear regime $\xia \gg 1$ (Balian \& Schaeffer 1989a). At very large overdensities one gets a pure exponential cutoff:
\beq
\rhot \gg \xia \; : \;\;  P_l(\rhot)  \simeq  \frac{a_{s,l}}{\xia^{\;2}} \left( \frac{\rhot}{\xia} \right)^{\omega_{s,l}-1}  \exp \left[ - \frac{\rhot}{x_{s,l} \; \xia} \right]  
\label{hl}
\eeq
In the intermediate range $P_l(\rhot)$ is a power-law:
\beq
\xia^{\; \frac{-\omega_l}{1-\omega_l}} \ll \rhot \ll \xia \; : \;\;\;   P_l(\rhot)  \simeq  \frac{a_l (1-\omega_l)}{\Gamma(\omega_l) \xia^{\;2}} \; \left( \frac{\;\rhot\;}{\;\xia\;} \right)^{\omega_l-2}   
\label{hs}
\eeq
At very low densities (``voids'') one obtains an exponential cutoff:
\beq
\begin{array}{l} {\displaystyle P(\rhot)  \simeq   a_l^{\frac{-1}{1-\omega_l}} \; \xia^{\; \frac{\omega_l}{1-\omega_l}} \; \sqrt{ \frac{(1-\omega_l)^{1/\omega_l}}{2 \pi \omega_l z^{(1+\omega_l)/\omega_l} } } } \\ \\  {\displaystyle \hspace{2.5cm} \times \exp \left[ - \omega_l \left( \frac{z}{1-\omega_l} \right)^{-\frac{1-\omega_l}{\omega_l}} \right] } \end{array}
\label{Pz}
\eeq
\[
\mbox{if} \hspace{0.3cm}  \rhot \ll \xia^{\; -\omega_l/(1-\omega_l)} 
\]
where we defined $z = \rhot \; a_l^{-1/(1-\omega_l)} \; \xia^{\; \omega_l/(1-\omega_l)}$. This behaviour of the probability distribution $P_l(\rhot)$ is shown in Fig.\ref{figPrho}.

Of course, since $\varphi_l(y)$ defines $P_l(\rhot)$ it also determines the moments $\mu_p$ (see Balian \& Schaeffer 1989a, 1989b, for details). Using (\ref{Pl}) one obtains:
\beq
\mu_p = \Gamma(p) \; \xia^{\;p-1} \; \inta \frac{dy}{2\pi i} \; (-y)^{-p} \; \varphi_l'(y) \; e^{-\varphi_l(y)/\xia} 
\eeq
with Re$(y)<0$. For large $p$ and $\xia$ one can drop the exponential:
\beq
\begin{array}{l} {\displaystyle p > (1-\om_l) \; , \; \xia \gg 1 \; : } \\ \\ {\displaystyle  \hspace{1cm} \mu_p \sim \Gamma(p) \; \xia^{\;p-1} \; \inta \frac{dy}{2\pi i} \; (-y)^{-p} \; \varphi_l'(y) } \end{array}
\eeq
In particular, for $p \in \naturs$ we recover (\ref{mupxip}):
\beq
p \in \naturs \; , \; \xia \gg 1 \; : \hspace{0.3cm} \mu_p(l) \sim S_p(l) \; \xia^{\;p-1} = \xia_p(l) 
\eeq
On the other hand, we can also write from (\ref{Pl}):
\beq 
\mu_p = \frac{\xia^{\;p-1}}{\Gamma(1-p)} \; \int_0^{\infty} dy \; y^{-p} \; \varphi_l'(y) \; e^{-\varphi_l(y)/\xia}
\eeq
which leads to:
\beq
\begin{array}{l} {\displaystyle p < 1-\om_l \; , \; \xia \gg 1 \; : } \\ \\ {\displaystyle  \hspace{1cm} \mu_p \sim \frac{\Gamma \left( 1-\frac{p}{1-\om_l} \right)}{\Gamma(1-p)} \; a_l^{p/(1-\om_l)} \; \xia^{\;-p \om_l/(1-\om_l)} } \end{array}
\label{muom}
\eeq
Note that (\ref{muom}) gives the exact leading order in the limit $p \rightarrow -\infty$ at a fixed value of $\xia$, or in the limit $\xia \rightarrow \infty$ at fixed $p$.

\end{document}